# An Architecture for Integrated Intelligence in Urban Management using Cloud Computing

Zaheer Khan, David Ludlow and Richard McClatchey
University of the West of England,
BS16 1QY, Bristol, United Kingdom
{Zaheer2.Khan, David.Ludlow,
Richard.McClatchey}@uwe.ac.uk

Ashiq Anjum
School of Computing and Mathematics
University of Derby, Kedleston Road,
DE22 1GB, Derby, United Kingdom
A.Anjum@derby.ac.uk

*Abstract*— **With the emergence of new methodologies and technologies it has now become possible to manage large amounts of environmental sensing data and apply new integrated computing models to acquire information intelligence. This paper advocates the application of cloud capacity to support the information, communication and decision making needs of a wide variety of stakeholders in the complex business of the management of urban and regional development. The complexity lies in the interactions and impacts embodied in the concept of the urban-ecosystem at various governance levels. This highlights the need for more effective integrated environmental management systems. This paper offers a user-orientated approach based on requirements for an effective management of the urban-ecosystem and the potential contributions that can be supported by the cloud computing community. Furthermore, the commonality of the influence of the drivers of change at the urban level offers the opportunity for the cloud computing community to develop generic solutions that can serve the needs of hundreds of cities from Europe and indeed globally.**

## I. INTRODUCTION

Integrated environmental intelligence can be described as the capability of a system to access, process, visualise and share data, metadata and models from various domains (such as land-use/cover, biodiversity, atmosphere and socio-economic) for various purposes. Examples include the identification of environmental changes and their causality, the identification of hotspots such as renewable energy sources, studies of urban sprawl, future environmental trend analysis, socio-economic development, policy development and collaborative decision-making. However, environmental data is continuously increasing and is mostly fragmented, un-harmonised, it exists in proprietary and open systems, it is less compliant to standards and sometimes requires extensive computing capacity, which makes it difficult for it to be utilised across the platforms. This suggests that integrated environmental monitoring requires compliance to standards, data harmonisation and service interoperability together with extensive on-demand processing and storage capacities in order to answer science and policy related questions.

In the above context, this research is an attempt to bridge the gap between fragmented cross-thematic environmental information (in particular urban-ecosystems) from various local, regional and national sources. In this regard, it aims to develop a cloud-based framework that enables data accessibility and storage across the platforms, and provides necessary on-demand computational resources for necessary processing, simulations and visualisation tasks. This framework also intends to provide mechanisms for integrated information intelligence that can more effectively solve the expected and unknown environmental challenges such as climate change adaptation and/or mitigation. More specifically, from a user perspective, the aims are:
- to inform policy and science related complex questions which cannot be answered directly from disintegrated information sources;
- to study different environmental phenomena by investigating the causality relationship between various environmental variables;
- to enable access to cross-thematic environmental information for climate change mitigation and/or adaptation;
- to support various stakeholders in generating outputs for various environmental indicators targeting decision-making and policy implementing;
- to facilitate citizens in participatory environmental data collection and collaborative decision making processes; and
- to design and utilise state of the art user-centred information technology tools and techniques.

From a technological perspective, the above aims help in identifying the required system capabilities which mainly necessitate acquisition and harmonising cross-thematic spatial and non-spatial data (using standards) in order to provide a uniform access to interoperable information for various purposes. Further, it indicates that in order to deliver integrated intelligence from the increasing volume of environmental data, significant storage capacity to preserve increasing environmental data and computational power to process the data is needed. Also, it suggests that specialized processing and visualisation services are needed to analyse the data for better decision-making. For instance, when the output from processed data e.g. information maps, scale from a local (i.e. City) to national or regional levels for comparative environmental indicator analyses, the amounts of data and the number of processing requirements potentially increase. In this case, the cloud computing paradigm suits well to fulfill the above requirements. It can provide an on-demand framework to make systematic the information flow from data collection to analysis, monitoring and assessment based on standard guidelines such as ISO 19100 Series, the INfrastructure for SPatial InfoRmation in

Europe (INSPIRE) [1] guidelines and the Open Geospatial Consortium (OGC) [2] standards. Figure 1 depicts both the user and technological perspectives to achieve integrated information intelligence for environmental monitoring objectives. The challenge here is to introduce appropriate mapping, harmonisation, integration and utilisation of suitable tools and technologies in cloud environment in order to fulfill stakeholder requirements.

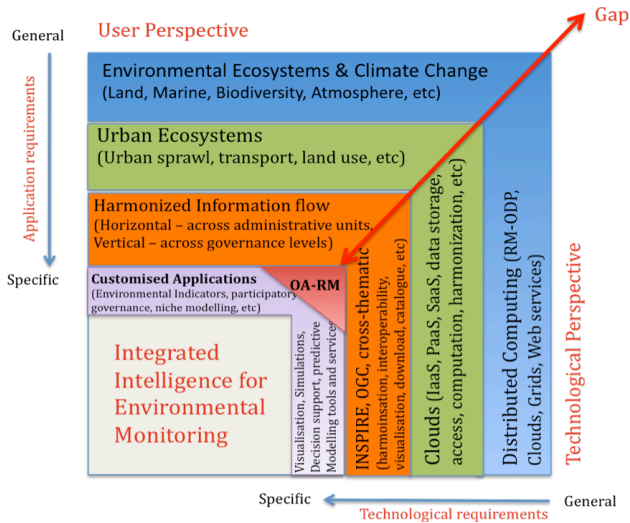

Figure 1: User-oriented and Technological-based perspective

The remainder of this paper is structured as follows. In section II, background and related work are presented. In section III, a user perspective is presented briefly in order to derive technological capabilities in clouds as presented in section IV. We present a proposed architecture for Integrated Environment Monitoring (IEM) in section V, followed by an assessment of the proposed architecture by identifying its benefits, potential research and technological challenges and an overall reflection on the expected research outcomes in section VI. Finally, we conclude in section VII.

## II. BACKGROUND AND RELATED WORK

The development of data harmonisation tools, decision-support systems, visualisation and simulation tools for specific environmental thematic domains is challenging. But it is not as challenging as the development and implementation of an integrated intelligence framework that can accommodate cross-thematic data harmonisation for various application domains using variety of tools, components and services.

In the above context, the Global Earth Observation System of Systems (GEOSS) [3] and Ground European Network for Earth Science Interoperations-Digital Earth Communities (GENESI-DEC) [4] are examples of ongoing initiative for integrating domain specific environmental data from various sources. More specifically, recent development of new methodologies for information acquisition (for example, the Shared Environmental Information System (SEIS) [5], Global Monitoring for Environment and Security (GMES) [6], and information provided by member states to European Environment Agency (EEA) due to different Directives such as Air quality, Noise and Water) and increase of computing capacities (e.g. sensor nets, smart phones, Web 2.0, grids and clouds) opens new fields for monitoring and assessment with high information flow and capacity to manipulate it.

From a technical perspective, managing the increasing data volumes, the varying needs of scalable computing resources, harmonisation and interoperability processes, tools and services, visualisation and compliance to standards are the major elements to develop an integrated environmental information system. In this regard, cloud computing paradigm is a suitable cyber-infrastructure to fulfill most of the above requirements due to its various characteristics [7]. The scope of cloud computing may vary from a private (e.g. for a single organization use) to public (e.g. access provided beyond an organization's use) or a combination of both (also called hybrid clouds). The scale of service provision also varies from Infrastructure as a Service (IaaS) to Platform as a Service (PaaS) to Software as a Service (SaaS). There are many commercial vendors already providing public cloud services: for example, Microsoft (Azure), Amazon (EC2/S3), Google (App Engine) and Rackspace etc. Similarly, some well-known research based initiatives are also under development e.g. Eucalyptus, OpenNebula, Nimbus, OpenStack.

There exist cloud (and in particular SaaS) enabled tools for different environmental applications such as land, transport and urban planning, e.g. ESS [8]. Similarly, the UK Natural Environmental Research Council has funded a research project called Environmental Virtual Observatory pilot (EVOp) [9] to use clouds to achieve similar objectives for the soil and water domains. However, further steps are needed to realise cross-thematic integrated intelligence for environmental ecosystems, using standard harmonisation processes in order to facilitate information requirements at all levels of governance i.e. local, national, and EU.

## III. AN EXAMPLE: USER PERSPECTIVE FOR AN URBAN ECOSYSTEM

At the local level, agencies addressing the management of urban regions have a critical need to integrate large sets of data across the sectoral domains of land-use, transport, health, etc in order to respond to the political demands for economically viable, socially cohesive and environmentally sustainable urban environments. The degree of complexity of an integrated system increases further when territorial cross-border aspects are also considered such as information based on the data from two or more cities (i.e. inter-city), two or more regions (i.e. inter-region) or two or more member states. For example, in Figure 2 some generic requirements have been indicated to capture typical integration challenges. It also shows that in this process of connectivity active communication and data integration between the levels of governance from local to EU is vital. Accordingly a model of both horizontal and vertical integration of other domain specific but related domains is defined in which the integration of necessary data flows is a prime objective.

Necessarily, urban monitoring aims to respond to the key political concerns of European cities with climate change, greenhouse gas emissions, uncontrolled urban sprawl, urban health and biodiversity loss etc, all of which are fundamentally related in the land-use-transport-environment nexus. Managing risk is far from straightforward, as cities are extremely complex systems, and the various drivers of change, impacts and responses are strongly interrelated, and can support, alter or compete with each other. Failure to integrate policy can be attributed to a variety of factors including notably organisational and procedural barriers between horizontal sectoral responsibilities for land-use, transport and environment, primarily at the local level, and between agencies responsible for policy development at local, regional, national and EU levels.

These failures underpin the concerns of the policy-making community for integrated urban management frameworks and are reflected in the demand for new assessment methodologies for urban and regional development. For instance, EEA's Integrated Urban Monitoring for Europe (IUME) initiative [10] connects the key political drive for climate change mitigation and adaptation, with associated priorities to ensure healthy and economically viable urban communities.

Improved integrated intelligence offers a major opportunity to address and overcome these deficiencies in policy responses necessary to secure the outcomes that combine the delivery of sustainable urban development or smart urban ecosystems and climate change amelioration. This also requires stakeholder engagement and necessitates new mechanisms for building the capacity and ability of community for creating sustainable futures, as advocated by Yigitcanlar T. [11], by introducing citizen science, participatory governance and collaborative decision-making.

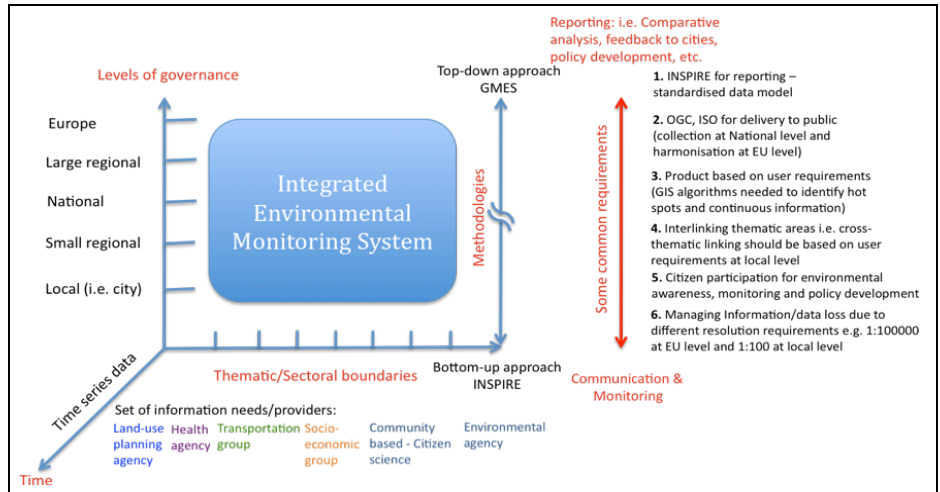

Figure 2: Integrated Environmental Monitoring: Needs and Challenges

## IV. MAPPING THE USER-ORIENTED PERSPECTIVE TO TECHNOLOGICAL PERSPECTIVE

Using the above user perspective here we attempt to present a common process that is used for environmental management with the objective of identifying technical capabilities required for the implementation of an IEM System (IEMS).

### A. A Generic Process for Environmental Management Using ICT

In the following we attempt to present a general monitoring process with the objective of mapping process activities to specific methods, technologies and tools for integrated environmental information intelligence. The main steps of the process are:

***Step 1: Data Acquisition:*** several approaches can be considered for this step including automated discovery and retrieval from distributed repositories, remote sensing, participatory sensing and citizens' science, etc. For example, EEA requires member-states to produce domain specific (i.e. air quality, noise, water) harmonised catalogues, a meta-database including data sources, data sets and other information such as reports and published meta information via EEA guideline compliant national web sites. Local governments are also active in local environmental monitoring initiatives e.g. SI@M in Vitoria-Gasteiz [12]**.**

***Step 2: Data Assimilation:*** this step is mainly needed to ensure that captured data meets standard schema structures and may require the harmonisation and transformation of the data according to required target system specifications.

***Step 3: Preliminary Data Analysis:*** this step is performed to determine whether there is need for additional data for required processing services.

***Step 4. Data Integration and Processing:*** this includes discovery, storage and integration of the required datasets.

***Step 5. Application Specific Outputs:*** this includes processing for various environmental models and visualisation of results for decision-making, etc.

**Step 6. Data Export:** this step is needed for further processing, storage and use in different tools and applications such as visualisation needs.

### B. Required Technical Capabilities for an IEM System

In general, the above generic process indicates that the following capabilities should be considered as technical requirements for integrated intelligence in an integrated environmental monitoring system:

*1) Data Acquisition*: the ability to collect data from various sources including databases, flat files, web services, sensors networks and web portals (e.g. OGC's Sensor Observation Service and Web Feature Service), participatory sensing and citizens' observations (e.g. using smart phones). This would further require:

  *a) Discovery:* the ability to discover relevant (meta-)data from various sources that match specific criteria, for instance, the OGC discovery service standard. Also use of common vocabularies or ontologies to mask semantic heterogeneity can be useful.

  *b) Metadata:* the ability to get additional information about the data such as ownership, usage restrictions, cost models, etc and compliance to metadata standards such as Dublin Core and ISO 19115.

  *c) Data quality:* the ability to assess the quality of retrieved data, e.g. missing data, resolution support, etc.

  *d) Retrieval:* the ability to retrieve the required data from a specified source using various methods such as OGC download service standard, File Transfer Protocol (FTP), Hyper Text Transfer Protocol (HTTP), eXtensible Markup Language-Remote Procedure Call (XML-RPC), etc.

  *e) Storage:* the ability to store transition and processed data.

*2) Schema mapping and transformations*: the ability to perform schema mappings to reference data sets, to harmonise spatial schema based on ISO 19100 series and INSPIRE specifications, for instance, coordinate transformation using different coordinate reference systems.

*3) Service interoperability*: the adoption of standards such as W3C's web standards, OASIS's RM-ODP [13] and OGC's view, download, discovery, catalogue services.

*4) Data fusion, processing and synthesis:* the ability to integrate data and apply computational processing steps e.g. using OGC WPS standard, in order to generate desired results and build synthesis around gaps in data coverage.

*5) Workflow management:* the ability to design, compose and execute workflows, e.g. using Kepler, Taverna, LONI.

*6) Provenance:* the ability to preserve and track information about sources of data and processes.

*7) Visualisation:* the ability to generate data and processing outputs in an user-friendly i.e. human understandable way by using various GUI techniques (e.g. OpenGL standard), 2D/3D maps, simulations, gaming, etc.

*8) Decision-making:* the ability to enable users to take decisions based on the recommended 'best-fit' output from various scenarios using artificial intelligence (predicate, description, fuzzy logic), expert systems tools and techniques (DROOLS and JESS rule-based engines).

*9) Social-networking:* the ability to enable users to interact with each other and share experiences (e.g. Web 2.0, forums, etc).

*10) Feedback mechanisms:* the ability to enable users to provide feedback/comments on results, annotate data and processing outputs (e.g. Web 2.0, Wikis, Issue Tracking).

*11) Security and reliability:* the ability to implement authentication, authorisation, encryption, decryption, auditing and backup mechanisms to enable use of data and services by legitimate users and avoid loss of data.

*12) Extensibility:* the ability to add new users, new data sources and new application-specific models, etc.

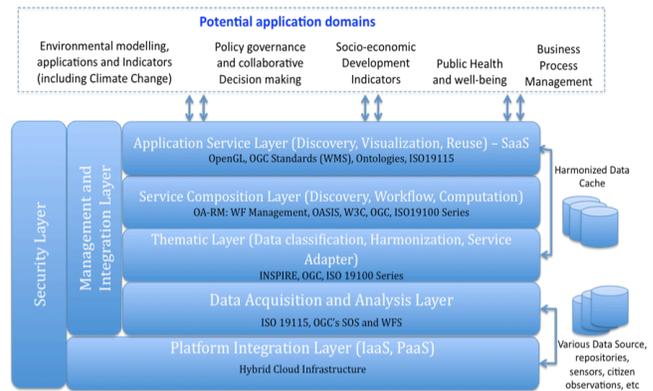

Figure 3: Proposed Architecture of IEMS

## V. PROPOSED ARCHITECTURE FOR IEM SYSTEM

Figure 3 depicts a proposed architecture from a technological perspective. It mainly consists of five horizontal and two vertical layers. The output from the first two bottom layers is generic which can be tailored to specific application needs in the above three layers.

- The *platform integration* layer depicts a cyber-infrastructure based on a hybrid cloud environment that ensures cross-platform accessibility of environmental data.
- The *data acquisition and analysis* layer is used to access environmental data from various sources including remote database repositories, sensor nets, citizens' observations in the cloud environment. This layer also ensures the quality of data acquired and identifies the need for necessary data harmonisation and data cleansing.

- The *thematic* layer classifies the acquired data into application specific thematic categories and performs data harmonisation and updates the data/service catalogues for further use of the data.
- The *service composition* layer is needed to design workflows, identify data sources, and link necessary processing components to enact the workflows. Furthermore, necessary analytical analysis of the workflow outputs can be performed at this layer. This layer also ensures that the provenance of data and specific processes is maintained that can be utilised for analysis by different expert systems at the *application* layer.
- The *application service* layer uses the outcomes from the service composition layer in application domain specific tools such as simulations and visual maps to perform analytical analysis for decision-making. Further, this layer enables stakeholders to use existing tools and develop new application domain specific components and services.
- The *management and integration* layer is used to automate the flow of data and information between the horizontal layers. It ensures that processed outputs from one layer to other are syntactically correct. It also aims to handle change management that occurs at different layers and intends to lessen the extent to which layered architecture requires management overhead.
- The *security* layer ensures necessary authentication, authorization and auditing for the use of data and services by legitimate users.

## VI. Assessment and Reflection on the IEMS Architecture

The IEMS layered architecture has the following benefits but it reveals several core challenges which require further investigation. In addition, we critically reflect on the extent to which this IEMS architecture is suitable for the stated objectives in section 1.

### A. Benefits of the Layered IEMS Architecture

The layered design of the architecture adopts the 'separation of concerns' as a core design principle in order to avoid reinventing the wheel and to capitalize on the experiences and knowledge from existing projects, tools and services. The layered architecture also brings extensibility and flexibility as new components can be added at various layers. Likewise outputs from different layers can be tailored to the specification of various tools and utilised in different applications.

For example, the architectural design of the IEMS aims to adopt ORCHESTRA Architecture Reference Model (OA-RM) [14] based approach in a cloud environment since it can provide a useful specification for data and service interoperability in a distributed environment. Similarly, for spatial data harmonisation there exist several tools and services which can be reused for the above purpose, e.g, the HUMBOLDT project [15] provides various INSPIRE and OGC compliant state-of-the-art harmonisation tools and services.

### B. Technological Challenges

The implementation of the IEMS architecture is not straightforward since it poses the following challenges:

*1)* The layered architecture provides flexibility in managing functionalities at individual layers. However, managing the complexity of various layers and integrating outputs from one layer to other in a (semi-)automated process chain is not straightforward and requires rigorous management of tasks at various layers. Therefore, a management and integration layer must be introduced to handle this complexity.

*2)* To provide a virtual environment for various stakeholders e.g. citizens, research, scientific and policy making communities, enabling them to select multiple variables across various environmental ecosystems (e.g. urban, transport, biodiversity and health) and to extract corresponding harmonised information from a cloud infrastructure which could be used for various purposes such as modelling, simulations, environmental indicators and decision making for policy development. In this regard, access to environmental data from different distributed repositories is needed. However, these distributed data sources may have different technological infrastructures including proprietary systems, desktop based systems, web services, grids, clouds, and non-standardized data models, which makes this task more challenging.

*3)* To adopt the uptake of cross-thematic harmonisation of selected environmental thematic areas such as land-use, transport and health according to existing standards such as INSPIRE Directive. The implementation of a thematic layer (Figure 3) in a cloud infrastructure based on INSPIRE components including implementation rules, metadata, data specification, services, monitoring and reporting will be challenging due to the recency of INSPIRE Annexes.

*4)* To enable various stakeholders to visually design workflows (e.g. using BPMN), to compose and bind services for execution, to identify data sources and to execute these workflows in a cloud environment. In addition, this requires the development of provenance mechanisms to keep track of data and service sources and intelligent processing performed for future references.

*5)* To develop case studies around citizen science i.e. to enable citizens to participate in environmental monitoring in order to raise their awareness and responsiveness towards environmental ecosystems using tools such as smart phones applications (for participatory sensing) and Web 2.0 (for collaborative decision-making). This would also require acquisition of data using standard interfaces such as OGC's Sensor Observation Service (SOS) and OGC's WFS and visualisation using OGC's WMS in a cloud environment.

*6)* Data may originate from two different cloud infrastructures. However, each cloud infrastructure is unique and mostly incompatible with each other i.e. the underlying cloud architecture, data models and access mechanisms, and services vary from one cloud infrastructure to another. Such an incompatibility introduces a gap in acquiring integrated intelligence across the platforms not only for applications from the same domain but also for cross-disciplinary applications such as environment, health and business. The main challenge here is to test hybrid clouds for integrated information system by adopting INSPIRE and OGC standards. This inter-cloud cross-thematic environmental ecosystem aspect brings novelty and an unique perspective in cloud computing research and, at the same time, will benefit both scientific, policy making and business communities in exploiting the full potential of clouds.

## C. Reflection and Critical Analysis

The real significance of integrated information intelligence is to be able to answer science and policy questions which cannot be answered directly from fragmented information sources. For example, in relation to urban and regional governance, integrated intelligence offers to genuinely support evidence based practice and to provide urban planners with the tools and methodologies that will enable them to effectively manage the complexity of the city. This will enable different stakeholders (environmentalists, urban planners, policy makers and citizens), who cannot deal with the complexity of the increasing data, harmonisation, service interoperability and processing needs, to get benefit from this work.

The proposed IEMS architecture attempts to provide the generic required capabilities (as outlined in section IV) in order to fulfill the requirements for the development of an integrated intelligence system for environmental monitoring. Despite there being several tools which can provide specific technical capability, integrating and using these tools in a cloud environment in compliance to different standards is not straightforward and hence necessitates several technological challenges as discussed in section VI.

The benefit of a cloud environment is twofold. Firstly the use of SaaS in cloud environment encapsulates the complexity of data acquisition, cross-thematic harmonised transformations, computer intensive processing, multi-dimensional modeling and visualisation and collaborative decision support mechanisms for various stakeholders. Secondly the extensibility and scalability characteristics of cloud platforms will accommodate continuously increasing data volumes and caching for visualisation and user groups who can be involved for citizen science-based participatory environmental monitoring.

One of the critical aspects the above research attempts to cover is building the capacity and ability of community for creating sustainable futures, and to be involved in community driven environmental decision-making processes. For example, citizens can have access to informed environmental intelligence effecting citizens' daily lives and regional economic prosperity which is directly or indirectly affected by environmental well-being. Similarly, urban planners can obtain improved predictions of impacts of urban planning policy by incorporating multiple variables such as land-use, socio-economic, transportation, public health. Furthermore, the processed output can be presented in various forms that could be used to indicate effectiveness of environmental changes at various levels (local, national and European) with benefits and consequences to stakeholders, policy and decision-makers and general public. However, rigorous development efforts are needed from technological peers to develop solutions based on the identified technical capabilities and proposed IEM architecture to achieve above aims.

## VII. CONCLUSIONS

In this paper, we have argued that the integration of fragmented environmental information can result in better environmental monitoring in order to mitigate various environmental challenges such as climate change. Our urban-ecosystem based example helps in identifying a generic set of technical capabilities for information intelligence and proposed a layered architecture for IEMS using clouds. However, it is not straightforward to realise integrated intelligence to its full potential due to certain technological challenges. These challenges require rigorous investigation from technological experts as future work in realising the true potential of the proposed architecture.